\documentclass[prl,twocolumn,showpacs,preprintnumbers,amsmath,amssymb]{revtex4}

\usepackage{graphicx}
\usepackage{dcolumn}
\usepackage{bm}

\begin{document}

\title{Slip line growth as a critical phenomenon}

\author{Fabio Leoni}
 \affiliation{Dipartimento di Fisica, Sapienza Universit\`a di
 Roma, P.le A. Moro 2, 00185 Roma, Italy.}
\author{Stefano Zapperi}
\affiliation{ CNR-INFM, S3, Dipartimento di Fisica, Universit\`a di Modena e Reggio
Emilia, via Campi 213/A, I-41100, Modena, Italy\\
ISI Foundation Viale San Severo 65, 10133 Torino, Italy
}

\begin{abstract}
We study the growth of slip line in a plastically deforming crystal 
by numerical simulation of a double-ended pile-up model with a
dislocation source at one end, and an absorbing wall at the other end.
In presence of defects, the pile-up undergoes 
a second order non-equilibrium phase transition as a function of stress, 
which can be characterized by finite size scaling. We obtain a complete
set of critical exponents and scaling functions that describe the
spatiotemporal dynamics of the slip line. Our findings
allow to reinterpret earlier experiments on slip line kinematography
as evidence of a dynamic critical phenomenon.
\end{abstract}

\maketitle

Plastic deformation in crystals is due to the motion of dislocations
driven by the external stress. In recent years, experimental and theoretical 
work has shown that dislocation dynamics is a complex intermittent phenomenon
involving the collective motion of many interacting dislocations \cite{zaiser06}.
In particular, deformation test on micron-scale crystals has revealed intriguing size effects
and power law distributed strain bursts \cite{uchic04,dimiduk06}, which have been 
reproduced in dislocation dynamics simulations \cite{csikor07}. Intermittent
dislocation motion is a generic feature of plasticity, not only in micron-scale  
samples, as shown recently by acoustic emission measurements \cite{weiss97,miguel01}
and by earlier reports from slip line kinematography in macroscopic samples 
\cite{tinder73,pothoff83,godon84}. 
These experimental observation lead to
the idea that plastic yielding is a non-equilibrium critical point \cite{zaiser06}, similar
to the jamming transition observed in soft and glassy materials  \cite{liu98}
or the depinning transition for disordered elastic manifolds \cite{kardar98}.

The yielding transition has been investigated in various dislocation models, 
from the dynamics of an individual flexible dislocation interacting with 
quenched random impurities such as solute atoms \cite{nabarro82,zapperi01}, 
to the dynamics of several rigid dislocations moving on single slip systems
\cite{miguel02,zaiser05}. While these models provided a good understanding 
of the yielding transition in simplified conditions, less is known about the role 
of dislocation nucleation and multiplication for the critical
behavior. In very clean single crystals, with a very small initial dislocation
density, dislocations are most likely nucleated from sources present at the surface
of the sample, where is easy to find defects (steps, scratches) acting as stress concentrators. 
In this case, the onset of plasticity corresponds to the creation and propagation of slip lines
through the entire cross section of the crystal. 

\begin{figure}
\begin{center}
\includegraphics[clip=true,width=8cm]{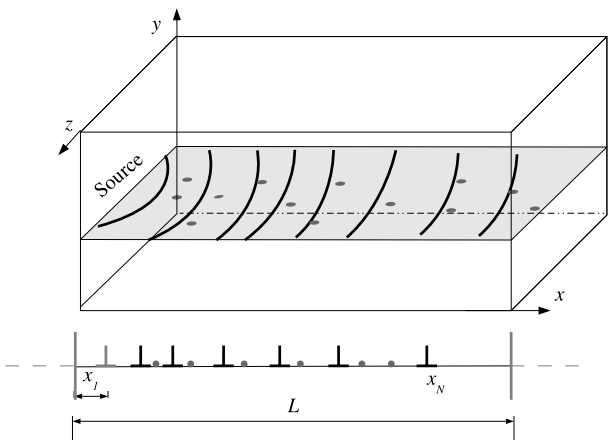}
\end{center}
\caption{\label{fig:1} At the top, we show a three dimensional sketch of a
  slip plane line containing a pile-up of dislocations emitted by a source
  placed on the left surface and absorbed at the right hand side of the
  sample. Our simplified model is shown in the bottom part of the figure:
  dislocations move on a line of length  $L$ and interact with point
  impurities. The source is modeled as an immobile dislocation with varying
  Burgers vector placed at $x_1$. In order to take into account the effect of
  open boundary conditions, interaction stresses include the contribution from
  an infinite series of image dislocations.}  
\end{figure}

A slip line can be envisaged as a queue of dislocations, a pile-up, pushed through a
series of obstacles, such as solute atoms or immobile dislocations from other
glide planes (see Fig.~\ref{fig:1}). Experimentally slip lines can terminate or
propagate depending on the value of the shear stress, temperature,
crystal structure and types of defects \cite{neuhauser83}.  A
transition from homogeneous to inhomogeneous slip, with increasing
impurity concentration, is observed experimentally in fcc alloys
\cite{neuhauser88,arkan87}. Here, we investigate the dynamics of a
double-ended pile-up in presence of defects, with a source of
dislocations at one end and an absorbing wall at the other. In the
model, the stress dependence of the stationary dislocation density,
velocity and strain rate obey finite-size scaling. This result
indicates that the transition observed in earlier experiments 
should be reinterpreted as a signature of a 
second-order non-equilibrium critical phenomenon. The
scaling exponents we measure in the present model are different from those found in the
corresponding homogeneous system, where nucleation is not considered
and the dislocation density is constant \cite{moretti04}.

We consider the pile-up as a group of identical edge or screw straight dislocations 
parallel to the $z$ axis that can move in the positive $x$ direction in the plane
$y=0$ when the net force acting on them is positive. This corresponds
to an effective one-dimensional model in which the dislocations are
generated from a source in the left side of a line of length $L$ (see Fig.~\ref{fig:1}).
They interact with each other and with a disordered stress
landscape provided by solute atoms, or other defects,
and disappear when they reach the right side of the line. The dislocations
have coordinates $x_i$, with $i=1,...N$, where $N=N(t)$ depends on
time $t$. The dislocation at $i=1$ is immobile and represents the source,
as we discuss below.  The dislocations for $i=2,...N$ are mobile and 
have constant Burgers vector $b_i=b$.  The Burgers vector is directed along 
$x$ for edge dislocations and along $z$ for
screw ones.

To describe the dynamics of mobile dislocations we use an over-damped 
equation, so that the velocity of dislocations depends linearly on the resolved
shear stress exerted on it \cite{kanninen69}. 
The equation of motion for the mobile dislocations is given by
\begin{equation}\label{i2N}
\chi\frac{dx_i}{dt} = b_i(\sigma+
\begin{array}{l}
\\
\mbox{\Large{$\sum$}}_{j=1}^{N}\\
\mbox{\scriptsize{$(j\ne i)$}}
\end{array}
\sigma_{i,j}^{int}+\sigma_i^{img})
+\sum_P f(x_i-X_P),
\end{equation}
where $\chi$ is an effective viscosity and $\sigma$ is the external stress.
The interaction stress $\sigma_{i,j}^{int}$ between dislocations $i$ and $j$ is
computed taking into account the image stresses of dislocation $j$ due to the open 
boundary conditions, while
$\sigma_i^{img}$ is due to the interaction between dislocation $i$ and its own images.
A compact expression of the interaction and image stresses can be obtained by performing
the sum over the images  \cite{hirth84}, yielding
\begin{equation}\label{eq:2}
\begin{array}{lll}
\sigma_{i,j}^{int} & = & -\dfrac{\pi}{2L}\dfrac{\mu b_j}{k}
\left[\cot\left(\pi\dfrac{x_j-x_i}{2L}\right)+\cot\left(\pi\dfrac{x_j+x_i}{2L}\right)\right],\\
&&\\
\sigma_i^{img} & = & \dfrac{\pi}{2L}\dfrac{\mu b_i}{k}
\cot\left(\pi\dfrac{x_i}{L}\right),
\end{array}
\end{equation}
where $\mu$ is the shear modulus, $k=\pi$ for screw dislocations and
$k=2\pi(1-\nu)$ for edge dislocations, $\nu$ is the Poisson ratio. We notice here, 
that the sum over the images is exact only in the case of screw
dislocations. For edge dislocations, there is 
an additional subdominant correction scaling as $1/r^2$ that we neglect here since
it should not influence the scaling behavior. 
The last term in Eq.~\ref{i2N} represents the interactions with pinning centers placed at 
randomly chosen positions $X_P$ with $P=1,...,N_P$. The detailed shape
$f(x)$ of the individual pinning force is inessential for most purposes, 
provided it is of short-range nature, and in this case it is given by
\begin{equation}
f(x)=-f_0\frac{x}{\xi_P}e^{-(x/\xi_P)^2},
\end{equation}
where $\xi_P$ is the range of the interaction and $f_0$ controls its strength.

Dislocations are typically generated by Frank-Read like sources, which
can only be represented in a three dimensional model. In lower dimensions, 
it is customary to model the source phenomenologically by creating dislocations
with a certain rate. The drawback of this approach is that the new dislocation
produces an artificial discontinuity in the stress field. To overcome this 
problem, we employ a method suggested by Zaiser \cite{zaiser-private} in 
which a source is represented by an immobile dislocation, placed at position $x_1$, 
with a time-dependent Burgers vector $b_1(t)$ growing with stress. 
When $b_1(t)=b$, a new mobile dislocation is emitted from the source whose
Burgers vector is reset to zero. The evolution equation for $b_1(t)$ is given by
\begin{equation}\label{b1_t}
\chi_1 \frac{db_1}{dt} = \theta(\sigma_1^{eff})\sigma_1^{eff} ,
\end{equation}
where $\theta$ is the Heaviside step function and $\chi_1$ is a damping constant
that we set equal to $\chi_1=\chi/b$. The effective stress
$\sigma_1^{eff}=\sigma+\sigma_1^{int}+\sigma_1^{img}$ is the sum of the constant
external stress $\sigma$, the stress $\sigma_1^{int}$ produced by the
interaction between the source and the mobile dislocations (including the
relative images), and the stress $\sigma_1^{img}$ produced by the interaction
between the source and its own images. These stresses are obtained
from Eq.~\ref{eq:2} observing that
$\sigma_1^{int}=\sum_{j=2}^{N}\sigma_{i=1,j}^{int}$ and
$\sigma_1^{img}=\sigma_{i=1}^{img}$.

Integrating numerically Eqs.~\ref{i2N} and \ref{b1_t}, we
analyze the dynamics of the pile-up as a function on the external stress
$\sigma$ and the system size $L$.
The units of time, space, and forces are chosen so that $b=1$, $\chi=1$ and $\mu/k=1$.
For the simulations reported here, we considered parameters 
$L=256,512,1024,2048,4096,8192$, $x_1=16$ and the pinning centers are Poisson distributed
with an average spacing $d_p=L/N_p=2$ with $f_0=1$ and $\xi=1$. 
From the experimental point of view, a key quantity describing 
the growth of the slip line is the 
plastic strain rate $\dot\epsilon$ given by
\begin{equation}
\dot\epsilon \equiv \frac{1}{L}\sum_i b_i\dot{x}_i = b\rho v,
\end{equation}
where $\rho=N/L$ is the dislocation density and $v=\sum \dot{x}_i/N$ is the average
dislocation velocity. Notice that all these quantities are defined
per unit dislocation length, given the effective one dimensional
geometry of our model. 

\begin{figure}
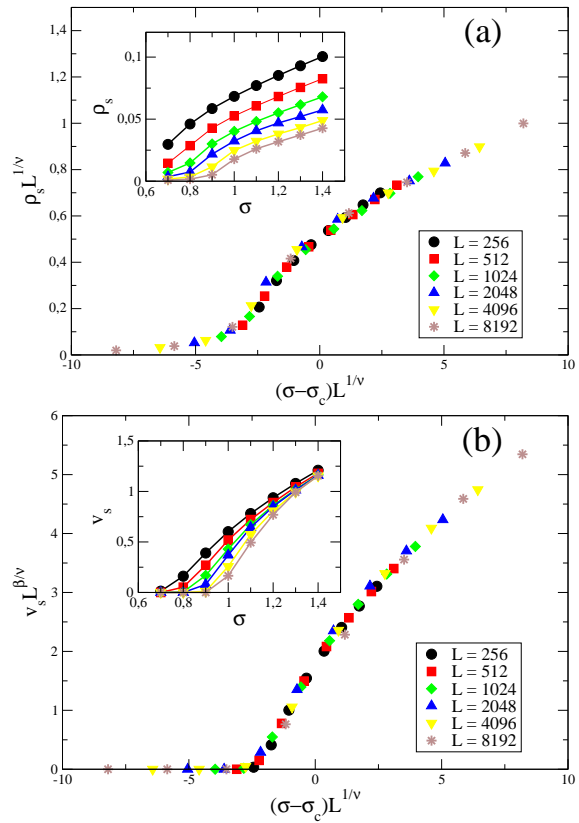

\begin{center}
\includegraphics[clip=true,width=7.5cm]{Fig2a}
\includegraphics[clip=true,width=7.5cm]{Fig2b}
\end{center}
\caption{\label{fig:2} (a) The average stationary density of the pile-up as a
function of the system size $L$ for different values of the applied stress
$\sigma$ (inset). The data collapse is consistent with the scaling hypothesis
in Eq.\ref{scaling} with the exponents $\alpha=1.00\pm 0.02$ and $1/\nu=0.35\pm 0.02$.
(b) The average stationary velocity of the pile-up as a
function of the system size $L$ for different values of the applied stress
$\sigma$ (inset). The data collapse is consistent with the scaling hypothesis
in Eq.\ref{scaling} with the exponents $\beta=0.48\pm 0.02$ and $\nu=2.85\pm
0.05$. } 
\end{figure}

Since the strain rate is simply the product of
the dislocation density and average velocity, we study directly 
these two quantities. We find that after an initial transient the
density and the velocity reach a steady state value ($\rho_s$ and
$v_s$ respectively) that depends on the system size $L$ and the applied stress
$\sigma$ as shown in the inset of Fig.~\ref{fig:2}. The graphs 
are suggestive of a non-equilibrium phase transition controlled 
by the stress between a pinned phase at low stress and a moving
phase at large stresses. The curves become sharper close to the
depinning point as $L$ is increased, as expected when 
finite-size effects are present. To confirm this idea we perform
a scaling collapse according to
\begin{equation}\label{scaling}
\begin{array}{lll}
\rho_s(\sigma,L) & = & L^{-\alpha/\nu}f[(\sigma-\sigma_c)L^{1/\nu}],\\
&&\\
v_s(\sigma,L) & = & L^{-\beta/\nu}g[(\sigma-\sigma_c)L^{1/\nu}],
\end{array}
\end{equation}
where the scaling function $f(u)$ fulfills the limits
\begin{equation}
f(u)\simeq\left\{\begin{array}{ll}
1 & \mbox{if $u\ll 1$},\\
u^{\alpha} & \mbox{if $u\gg 1$},
\end{array}\right.
\end{equation}
and for $g(u)$ they are 
\begin{equation}
g(u)\simeq\left\{\begin{array}{ll}
1 & \mbox{if $u\ll 1$},\\
u^{\beta} & \mbox{if $u\gg 1$}.
\end{array}\right.
\end{equation}
The best collapse is obtained using $\sigma_c=1.05\pm0.05$, $\nu=2.85\pm 0.05$, 
$\alpha/\nu=0.35\pm 0.02$ and $\beta/\nu=0.17\pm 0.02$, as shown in Fig.~\ref{fig:2}.
These exponent combinations correspond to $\alpha=1.00\pm 0.02$, $\beta=0.48\pm 0.02$,
and yield a scaling form for the strain rate of the type
\begin{equation}
\dot\epsilon(\sigma,L)= L^{-(\alpha+\beta)/\nu}h[(\sigma-\sigma_c)L^{1/\nu}],
\end{equation}
where $h(u)=f(u)g(u)$, as we have also verified directly. 

\begin{figure}
\begin{center}
\includegraphics[clip=true,width=7.5cm]{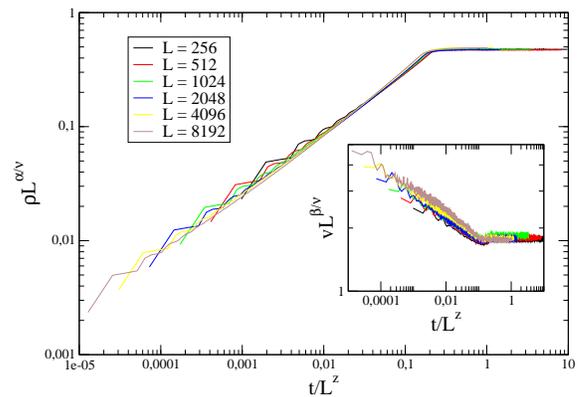}
\end{center}
\caption{\label{fig:3} The evolution of the dislocation density and velocity (inset) as a
function of the system size $L$ at the critical point $\sigma=\sigma_c=1.05$. 
The best collapse is obtained for $\alpha/\nu=0.35\pm0.02$, $\beta/\nu=0.17\pm0.02$
and $z=1.25\pm0.02$ which is consistent with the scaling collapse in Eq.~\ref{fig:2}.}
\end{figure}

We have also analyzed the slip line growth dynamics in the transient regime.
The time dependence of the dislocation density and velocity can also be characterized
by finite size scaling functions which at the critical point $\sigma=\sigma_c$
are given by
\begin{equation}
\begin{array}{lll}
\rho(t,L) & = & L^{-\alpha/\nu}f_t[t/L^z],\\
&&\\
v(t,L) & = & L^{-\beta/\nu}g_t[t/L^{z}],
\end{array}
\end{equation}
where $z$ the dynamic exponent. The best collapse is obtained
for $z=1.25\pm0.02$ as shown in Fig.~\ref{fig:3}. The scaling functions,
for small values of the argument, scale as $f_t(u) \sim u^\zeta$, with $\zeta=0.55\pm 0.05$
and $g_t(u)\sim u^{-\theta}$, with $\theta=0.10 \pm 0.05$. Hence, the strain rate 
in the initial phase grows as $\dot\epsilon(t) \sim t^{\zeta-\theta}$.
Notice that the scaling exponents are considerably different
from what is expected for a regularly spaced pileup with periodic boundary 
conditions, where the density is constant (hence $\alpha=0$) and  
the critical exponents are $\beta\simeq 0.78$, $z\simeq 0.78$ and $\nu\simeq 1.5$
\cite{moretti04}. 

\begin{figure}
\begin{center}
\includegraphics[clip=true,width=7.5cm]{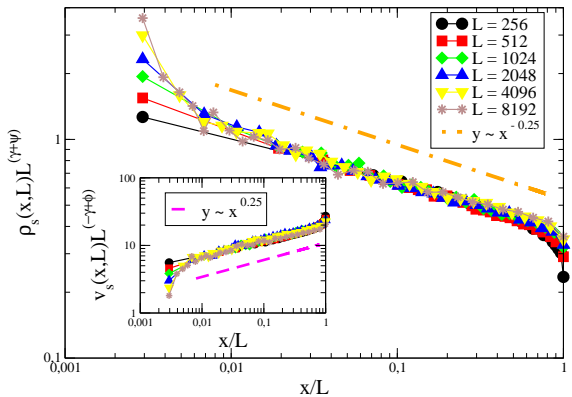}
\end{center}
\caption{\label{fig:4} The density $\rho_s(x,L)$ and velocity $v_s(x,L)$ (inset)
  profile for different values of the system size $L$, for $\sigma=1.05$ (critical
  regime). In the bulk of the line we have the power-law behavior
  $\rho_s(x,L)\sim x^{-\gamma}/L^{\psi}$ and 
  $v_s(x,L)\sim x^{\gamma}/L^{\phi}$ with $\gamma=0.25\pm 0.02$,
  $\psi=0.09\pm 0.02$ and $\phi=0.42\pm0.02$. These behaviors are
  consistent with the scaling of the stationary density and velocity
  in the critical regime.}    
\end{figure}

To elucidate the role of the boundary condition and 
characterize the internal morphology of the pile-up, we report in Fig.\ref{fig:4} the
stationary density $\rho_s(x,L)$ and velocity $v_s(x,L)$ profiles for
different values of the system size $L$, for $\sigma=\sigma_c=1.05$.  
We observe inhomogeneities for both density and velocity profiles which 
can be described as power laws: $\rho_s(x,L)\sim x^{-\gamma}/L^{\psi}$
and  $v_s(x,L)\sim x^{\gamma}/L^{\phi}$, with $\gamma\simeq
0.25\pm0.02$, $\psi\simeq 0.09\pm0.02$ and $\phi=0.42\pm0.02$. Notice
that the strain rate profile is approximately constant since 
the two power law cancel out in the product.
This behavior is consistent with the scaling of the stationary density and velocity
as described in Eq.~\ref{scaling}. In particular, we have that
\begin{equation}\label{profile_rho}
\rho_s(\sigma_c,L)=\frac{1}{L}\int_{x_1}^{L}\rho_s(x,L)dx
\sim L^{-(\gamma+\psi)}.
\end{equation}
Hence, we have the scaling relation $\gamma+\psi=\alpha/\nu$, that is verified by the
numerical values of the exponents. Similarly, from the steady-state
velocity equation
\begin{equation}
 v_s(\sigma_c,L) =  \frac{\int_{x_1}^{L}\rho_s(x,L)v_s(x,L)dx}{\int_{x_1}^{L}\rho_s(x,L)dx} \sim
L^{-(\phi-\gamma)}
\end{equation}
we obtain the relation $\phi-\gamma=\beta/\nu$, which is again 
in agreement with our numerical estimates.

In conclusion, we have studied the slip line formation at the initial stage of
plastic deformation in a crystal by means of the double-ended pile-up model 
finding that in presence of pinning centers (quenched disorder) the model
exhibit a non-equilibrium phase transition. As a consequence of this 
dislocation density, velocity and strain-rate are described by 
finite size scaling. Finite size scaling has direct implications for 
size effects: the size dependence of the yield stress $\sigma_Y$ observed in
micron scale plasticity \cite{uchic04}. Considering the
scaling law in Eq.~\ref{scaling}, we expect that the yield stress 
for finite $L$ grows towards the asymptotic value $\sigma_c$ according
to $\sigma_Y(L)=\sigma_c-A/L^{1/\nu}$, where $A$ is a positive constant. 
This type of inverse size effect, with the strength increasing with
the sample size, is due to the larger back-stress exerted on the source 
by the pile-up as its length is increased. In more general cases, involving many 
sources and several slip lines, the constant $A$ is expected to be negative 
as shown in other models of the yielding transition \cite{zaiser06}.

{\it Acknowledgments -} This work is supported by the European Commissions 
NEST Pathfinder programme TRIGS under contract NEST-2005-PATH-COM-043386.
We thank M. Zaiser for useful discussions.

\end{document}